\def\year{2022}\relax
\documentclass[letterpaper]{article} 
\usepackage{aaai22}  
\usepackage{times}  
\usepackage{helvet}  
\usepackage{courier}  
\usepackage[hyphens]{url}  
\usepackage{graphicx} 
\usepackage{natbib}  
\usepackage{caption} 
\DeclareCaptionStyle{ruled}{labelfont=normalfont,labelsep=colon,strut=off} 
\usepackage{algorithm}
\usepackage{algorithmic}
\usepackage{colortbl}
\usepackage[table]{xcolor}
\usepackage{amssymb}

\usepackage{newfloat}
\usepackage{listings}
\floatstyle{ruled}
\newfloat{listing}{tb}{lst}{}
\floatname{listing}{Listing}
\usepackage{multirow}
\usepackage{amsmath}

\title{RepBin: Constraint-based Graph Representation Learning for \\ Metagenomic Binning}

 \author {
     Hansheng Xue,\textsuperscript{\rm 1}
     Vijini Mallawaarachchi, \textsuperscript{\rm 1}
     Yujia Zhang, \textsuperscript{\rm 1}
     Vaibhav Rajan, \textsuperscript{\rm 2}
     Yu Lin \textsuperscript{\rm 1}\footnote{Corresponding author.}
 }
 \affiliations {
     \textsuperscript{\rm 1} School of Computing, The Australian National University, Canberra, Australia\\
     \textsuperscript{\rm 2} School of Computing, National University of Singapore, Singapore\\
     \{hansheng.xue, vijini.mallawaarachchi, yujia.zhang, yu.lin\}@anu.edu.au, 
     vaibhav.rajan@nus.edu.sg
 }

\begin{document}

\maketitle

\begin{abstract}

Mixed communities of organisms are found in many environments -- from the human gut to marine ecosystems -- and can have profound impact on human health and the environment.
Metagenomics studies the genomic material of such communities through high-throughput sequencing that yields DNA subsequences for subsequent analysis.
A fundamental problem in the standard workflow, called binning, is to discover clusters, of genomic subsequences, associated with the unknown constituent organisms.
Inherent noise in the subsequences, various biological constraints that need to be imposed on them and the skewed cluster size distribution exacerbate the difficulty of this unsupervised learning problem.
In this paper, we present a new formulation using a graph where the nodes are subsequences and edges represent homophily information.
In addition, we model biological constraints providing heterophilous signal about nodes that cannot be clustered together.
We solve the binning problem by developing new algorithms for (i) graph representation learning that preserves both homophily relations and heterophily constraints (ii)
constraint-based graph clustering method that addresses the problems of skewed cluster size distribution.
Extensive experiments, on real and synthetic datasets, demonstrate that our approach, called RepBin, outperforms
a wide variety of competing methods.
Our constraint-based graph representation learning and clustering methods, that may be useful in other domains as well, advance the state-of-the-art in both metagenomics binning and graph representation learning.


\end{abstract}

\section{Introduction}

The field of \textit{metagenomics} has paved the way to study entire microbial communities obtained from natural environments~\cite{kaeberlein2002isolating}. Large scale 
studies such as the Human Microbiome Project~\cite{turnbaugh2007human} have leveraged metagenomics analyses to gain valuable insights about the complex microbial communities found in the human body, and study the relationships of these microbial communities with health conditions and diseases such as pregnancy and preterm birth (PTB), inflammatory bowel diseases (IBD), stressors affecting prediabetes~\cite{integrative2019integrative}, influence of diet on metabolism~\cite{pasolli2020large,asnicar2021microbiome} and disease association of microbial species found in the human gut~\cite{nayfach2019new}.

In a typical metagenomics workflow, genetic material from a microbial community is collected and first sequenced using high-throughput sequencing (HTS) platforms.
The output at this stage contains short sequences of DNA called \textit{reads} of all the constituent microorganisms.
Note that since the genetic material of all the microorganisms are mixed to begin with, it is unknown which species each read belongs to; further, it is possible that multiple species may contain the same DNA sequence in their reads.
A fundamental problem, for any subsequent downstream analysis, is to identify the species involved in the input sample.


\begin{figure}[t]
\centering
\includegraphics[width=0.99\columnwidth]{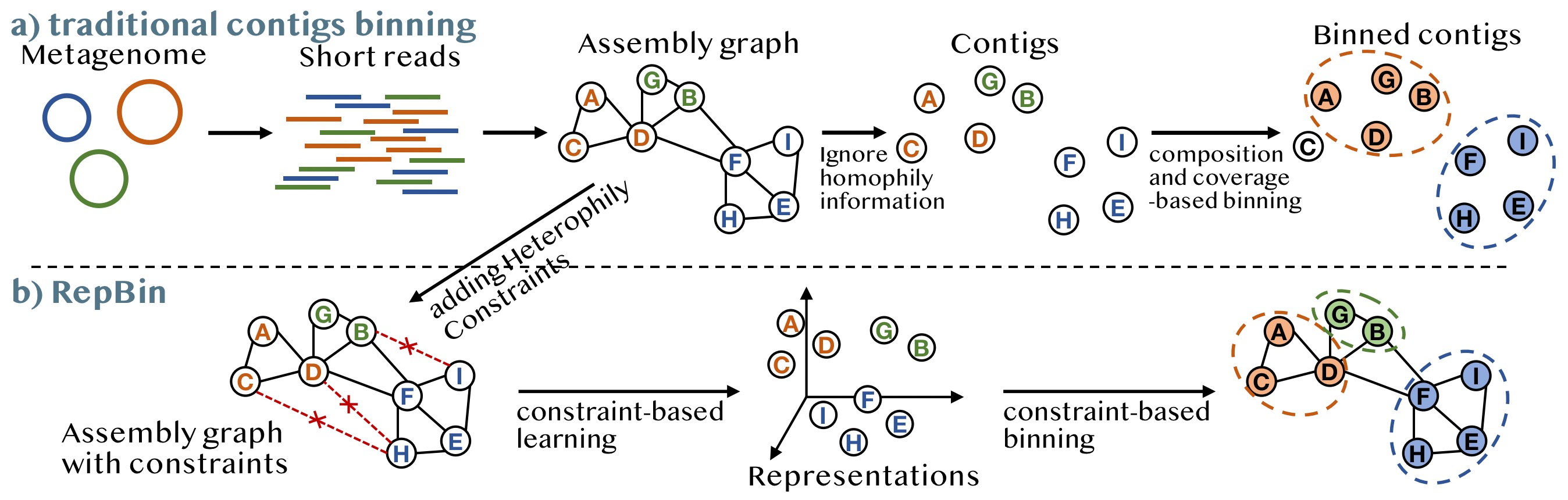}
\caption{The pipeline of traditional metagenomic contigs binning and our proposed method, RepBin.}
\label{fig1}
\end{figure}

Popular metagenomics approaches 
assemble these short reads into longer non-overlapping DNA sequences called \textit{contigs} using \textit{assembly graphs},  
where each vertex represents a contig and each edge represents the homophily information between two contigs~\cite{nurk2017metaspades}.
Then, 
these assembled contigs are clustered into specific bins corresponding to the constituent genomes in the microbial communities (referred as \textit{metagenomic binning}). 
Contigs connected to each other in the assembly graph are more likely to belong to the same species~\citep{barnum2018genome}.
Additional biological information can be used to incorporate constraints on contigs that are likely to be in different bins.
For instance, single-copy marker genes are those that appear just once in each species and so, if two contigs contain the same single-copy marker gene, they are most likely to belong to different species.
Such constraints can provide heterophilous information and could be utilized to improve binning.

Most metagenomics binning tools 
ignore the homophily information in assembly graphs and use the composition and coverage information of contigs for binning (refer to Figure~\ref{fig1}a for a traditional pipeline for binning). Moreover, these tools have to discard many short contigs and thus suffer from low recall values because the composition and coverage features become unreliable for short contigs.
To use the assembly graph directly, an alternative approach would be to obtain node embeddings~\cite{Cui2019ASO} from the assembly graph, and cluster them to find bins. 
However, existing graph embedding techniques mainly utilize homophily information and do not model heterophilous relations.


In this paper, we propose a constraint-based graph representation learning model, called \textit{Constraint-based Learning}, which can capture the structural information of the graph while respecting the prior constraints. The \textit{Constraint-based Learning} model is a contrastive graph learning framework with diffusion convolutional operators as basic components, and optimizes prior constraints through a penalty term in the objective function. The learned representations can be used for downstream tasks. For metaganomic contig binning, we propose a GCN-based label annotation model, called \textit{\textit{Constraint-based Binning}}, with constrained nodes as initial labels. The combined model is called \textit{\textbf{RepBin}} and is illustrated in 
Figure~\ref{fig1}b. 
Our contributions in this paper are: 
\begin{itemize}
    \item To the best of our knowledge, this is the first use of graph representation learning in the important area of metagenomic contig binning.
    \item We design a novel constraint-based graph representation learning model, called \textit{Constraint-based Learning}, which can capture structural information of the graph while respecting heterophilous constraints.
    \item We design a novel \textit{Constraint-based Binning} strategy which uses Graph Convolutional Networks to annotate  unknown contigs with labels, using constrained contigs to obtain initial labels.
    \item We benchmark the performance of our method on contigs binning, against state-of-the-art methods for metagenomics binning, graph representation learning and graph clustering. Results show that \textit{RepBin} significantly outperforms them 
    on both simulated and real-world datasets.
    \item Both algorithms, Constraint-based Learning and Binning, are general-purpose graph representation learning and clustering methods, and can be used in other domains as well, where constraints need to be incorporated in these tasks.
\end{itemize}

\section{Related Work}

\textbf{Metagenomic Binning.} 
Although contigs are assembled from short reads using assembly graphs in metagenomic samples, most existing binning tools ignore the homophily information in assembly graphs. Instead, these binning tools use the composition (normalised oligonucleotide, \emph{i.e.}, short string of length $k$, frequencies) and coverage (average number of reads aligning to each position of the contig) information to bin contigs, \emph{e.g.}, in MetaWatt~\cite{MetaWatt2012}, CONCOCT~\cite{alneberg2014binning}, and MaxBin2~\cite{Wu2015MaxBin2}. 
MetaBAT2~\citep{kang2019metabat} employs a graph clustering approach, where the graph is constructed by composition information of contigs. 
SolidBin~\citep{Wang2019SolidBin} uses semi-supervised spectral clustering that incorporates additional biological knowledge.
VAMB~\cite{nissen2021VAMB} uses deep variational autoencoders to integrate both composition and coverage information and clusters the resulting latent representation of contigs into bins. 
BusyBee Web~\citep{BusyBeeWeb2017} is a web-based application which makes use of a bootstrapped supervised binning approach to bin contigs. 
However, all these tools have to discard short contigs (\emph{e.g.}, shorter than 1000bp) 
because the composition and coverage features become unreliable for short contigs,
and thus suffer from low recall values. To recover these short contigs, recently published bin-refinement tools~\cite{vijini2020graphbin,mallawaarachchi2020graphbin2} have introduced the use of assembly graphs from which contigs are derived.

\noindent\textbf{Graph Representation Learning.} 
Existing unsupervised graph representation learning models are mainly grouped into three categories, random walk-based~\cite{Grover2016node2vec,Perozzi2014DeepWalk}, matrix factorization-based~\cite{qiu2018netMF,liu2019general}, and deep learning-based methods~\cite{kipf2016VGAE,velickovic2018DGI}. 
%
Generative and contrastive models are two typical frameworks for deep learning-based unsupervised graph learning models. 
Generative learning models, like VGAE~\cite{kipf2016VGAE}, capture the graph structure by minimizing the error between the constructed and original graph in an encoder-decoder architecture. Contrastive learning, such as DGI~\cite{velickovic2018DGI}, utilizes discriminator to differentiate nodes from input graph and negative samples. The node embeddings are optimized via maximizing mutual information with graph summary. 
%

\noindent\textbf{Graph Clustering}
Graph clustering aims to use the graph structure to group nodes in graph into several disjoint clusters, like spectral clustering~\cite{von2007sc}. 
Many deep graph clustering methods have been proposed~\cite{bianchi2020mincutpool,tsitsulin2020graph}. 
These methods mainly use deep learning or GNN models to capture the graph structure and then use clustering algorithms to cluster these learned features~\cite{cao2016DNGR,Fan2020O2MAC,Zhang2019AGC}. 
%
%
Methods such as
Constraint-based spectral clustering~\cite{wang2014constrained} and Deep constrained clustering~\cite{zhang2019dcc} 
are designed to integrate constraints in clustering.

\section{Preliminaries}
Consider an assembly graph $G=(V,E)$ from a metagenomic assembler, where each node $v \in V$ represents a contig and each edge $e \in E$ denotes that two corresponding contigs (nodes) are connected in the assembly graph. 
Let $\mathcal{M}$ be the set of all pairwise heterophily 
constraints induced by 
\textit{Single-copy marker genes} which are conserved and appear only once in most of the bacterial genomes~\cite{albertsen2013genome}. 
We use the tools FragGeneScan~\cite{rho2010fraggenescan} and HMMER~\cite{eddy2011accelerated} to identify contigs containing each of the 107 single-copy marker genes.
Then, we generate pairwise constraints between contigs containing each single-copy marker gene, \emph{e.g.}, there exists a constraint $m(u,v)$ between $u$ and $v$ if the same single-copy marker gene appears in both $u$ and $v$ (which indicates that $u$ and $v$ are not expected to be in the same species). 

The task of binning metagenomic contigs is to cluster contigs into bins that correspond to different species. Different from other models that directly use $k$-mer composition and contig coverage to bin contigs, we make use of the structure of assembly graph and label contigs based on the low-dimensional features learned from the assembly graph. The constraint-based representation 
of a contig from an assembly graph can be defined as:

\textbf{Definition.} Given an assembly graph $G=(V,E)$ and its 
constraints $\mathcal{M}$, the constraint-based embedding of a node (contig) in the assembly graph is a $d$-dimensional feature $H\in R^{|V|\times d}$, where $d\ll |V|$, that considers both the topological structure of the assembly graph and the constraints of $\mathcal{M}$ simultaneously.

\section{Methodology} 
\begin{figure}[t]
\centering
\includegraphics[width=0.99\columnwidth]{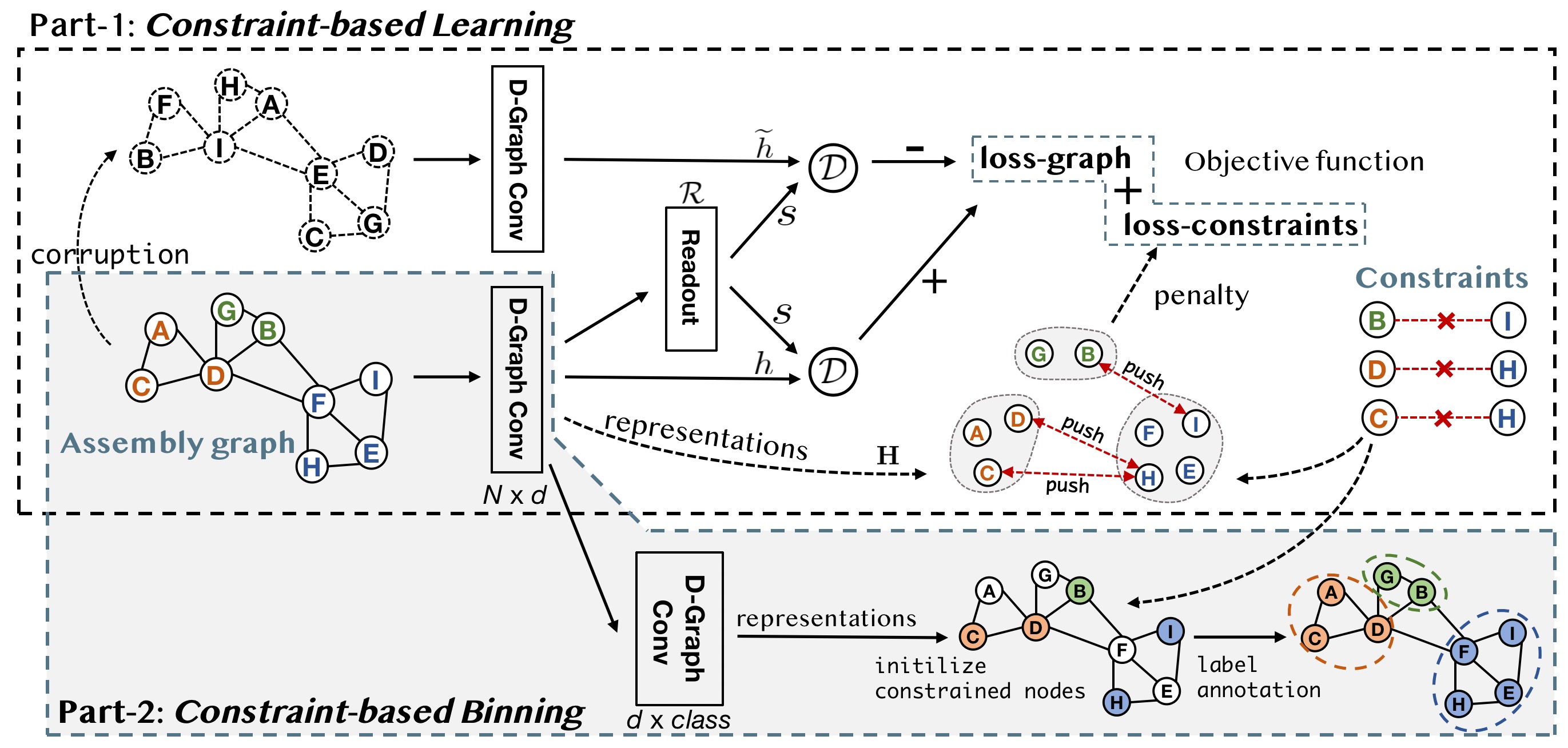}
\caption{The framework of our proposed RepBin model.}
\label{fig2}
\end{figure}

The proposed framework mainly contains two components, \textit{Constraint-based Learning} and \textit{Constraint-based Binning}. 
Figure~\ref{fig2} shows an overview of the entire RepBin model.

\subsection{Part 1: \textit{Constraint-based Learning}}
In the \textit{Constraint-based Learning} module, we aim to model the homophily graph structure and heterophily constraints. 
We first describe the contrastive graph learning framework and graph diffusion convolution, then explain how to integrate constraints to obtain the final node embedding matrix. 

\subsubsection{Contrastive graph learning framework.} 
Here we introduce a contrastive graph representation learning model which tries to obtain representations for nodes that capture the global structure of the entire graph by maximizing the mutual information between node-level (local) and graph-level (global) features. The overall framework of the \textit{Constraint-based Learning} is shown in the part 1 of Figure~\ref{fig2}, which is similar to DGI. 
The proposed model mainly contains graph encoder, negative graph sampling, readout function, discriminator model, and constraints optimization. 

The one-layer graph diffusion convolutional operator is used in the \textit{Constraint-based Learning} model as the graph encoder module. We will explain diffusion convolution module in the next subsection. By the diffusion convolution operator in Equation~\ref{eqn05}, we can obtain the representations of all nodes $\mathbf{H}=\{\mathbf{h_1},\mathbf{h_2}, ..., \mathbf{h_n}\}$. 



The negative graph sampling generates negative graphs by applying a corruption function (\emph{e.g.}, row-wise shuffling on the original attribute matrix, same as the one used in the DGI model) on the original assembly graph, $(\tilde{X},\tilde{A})=\mathcal{C}(X,A)$. 
Both the original and negative graphs are typed into the graph encoder model and generate latent node-level representations $\mathbf{H}$ and $\mathbf{\tilde{H}}$ respectively. 

The readout function $\mathbf{s}=\mathcal{R}(\mathbf{H})=\frac{1}{n}\sum_{i=1}^n\mathbf{h}_i$ is a function to obtain a global representation of the whole assembly graph by aggregating all node-level representations. 
The discriminator model $\mathcal{D}$ is introduced to discriminate the true samples, \emph{i.e.}, ($\mathbf{h_i},\mathbf{s}$), from its negative counterparts, \emph{i.e.}, ($\mathbf{\tilde{h_j}}, \mathbf{s}$), by maximizing the mutual information. 

Finally, we use a standard binary cross-entropy (BCE) loss between the samples from the joint (positive examples) and the product of marginals (negative examples) to optimize the mutual information (MI) between $\mathbf{h_i}$ and $\mathbf{s}$, same as DGI and DMGI~\cite{park2019DMGI}. 

\begin{equation}
\begin{aligned}
  \mathcal{L}_g=-\frac{1}{2n}[\sum_{i=1}^n\log\mathcal{D}(\mathbf{h_i},\mathbf{s})+\sum_{j=1}^n\log(1-\mathcal{D}(\mathbf{\tilde{h_j}},\mathbf{s}))]
\end{aligned}
\end{equation}

\subsubsection{Graph diffusion convolution.}
The main characteristic of the assembly graph which is different from arbitrary graphs is the high homophily score (see Table~\ref{tab:dataset}). 
It indicates that the majority of linked contigs belong to the same species. 
Most message-passing neural networks (\emph{i.e.}, GCN) mainly aggregate information from one-hop neighbors in each layer and are limited in exploring higher-order neighborhoods, especially in graphs with high homophily. 
In contrast, diffusion has been found to be superior in graphs with high homophily~\cite{klicpera2019GDC,klicpera_predict_2019PPNP}. So, in our model, we use the Graph Diffusion Convolution (GDC) operator to capture the high-order structure of the graph.




One successful example of the graph diffusion is the personalized PageRank (PPR)~\cite{page1999pagerank}. Assuming that one node $i$ and its teleport vector $x_i$, the adaptation of the PPR for node $i$ can be calculated by the recurrent equation $\textrm{PPR}(x_i)=(1-\alpha)\hat{\textbf{A}}\textrm{PPR}(x_i)+\alpha x_i$. Parameter $\alpha\in(0,1]$ denotes the probability of teleporting to another state. We can obtain the transitions state for node $i$ using $\textrm{PPR}(x_i)=\alpha(\textbf{I}_N-(1-\alpha)\hat{\textbf{A}})^{-1}x_i$. Thus, the normalized graph diffusion matrix for graph $G$ can be formulated as: 
\begin{equation}\label{eqn05}
\mathbf{T}_{S_{ij}} = \frac{\textbf{S}_{ij}}{\sum_i\textbf{S}_{ij}}\ 
\textrm{with}\quad \textbf{S}=\alpha[\mathbf{I}_N-(1-\alpha)\mathbf{T}]^{-1}
\end{equation}

where $\mathbf{T}=\mathbf{\hat{D}}^{-1/2}\mathbf{\hat{A}}\mathbf{\hat{D}}^{-1/2}$ is the symmetric transition matrix (approximately the spectral graph convolutional operator), $\mathbf{\hat{A}}$ is the adjacency matrix of the graph with self-loops $\mathbf{\hat{A}=A+I}_N$, $\mathbf{I}_N$ is the unit matrix, and $\mathbf{\hat{D}}$ denotes the diagonal degree matrix, \emph{i.e.}, $\mathbf{\hat{D}}_{ii}=\sum_{j=1}^N\mathbf{\hat{A}}_{ij}$. The calculated diffusion matrix is dense and computationally inefficient, we can optionally add a sparsity operator (truncating small values less than $\epsilon$: $\textbf{S}_{ij}(\epsilon) = \textbf{S}_{ij}\ \textrm{if}\ \textbf{S}_{ij} \geq \epsilon\ \textrm{else}\ 0$) before the normalization. 

We use graph diffusion convolution in the contrastive learning model to capture the graph structure. We formulate the $l$+1-layer diffusion convolutional operator following the layer-wise propagation rule as: 
\begin{equation}\label{eqn06}
    \mathbf{H}=\mathbf{\sigma}(\mathbf{T}_S\cdot\mathbf{X}\mathbf{\Theta})
\end{equation}
where $\sigma(\cdot)$ denotes a non-linear activation function  (\emph{e.g.}, ReLU), $\mathbf{\Theta}\in\mathbb{R}^{N \times d}$ is a trainable transformation matrix, $d$ is the feature dimension, and \textbf{X} is the initial feature matrix.


\subsubsection{Constraints optimization.}
In metagenomics, side information is given in an implicit way of negative constraints, which indicate pairwise contigs that belong to distinct species, and not in an explicit way (such as known labels for contigs). 
In our \textit{constraint-based learning} model, we 
penalize the constrained nodes 
with respect to their similarity as described below.

In the previous graph diffusion convolution operator, we can obtain latent representations for each node $\mathbf{H}=\{\mathbf{h_1}, \mathbf{h_2}, ..., \mathbf{h_n}\}$. These learned features capture the structural information of the assembly graph and ignore the pairwise heterophily constraints $\mathcal{M}$. 
Each pairwise constraint $m(i,j) \in \mathcal{M}$ indicates that node $i$ and $j$ should belong to two different species. We calculate the distance $(\mathbf{h}_i,\mathbf{h}_j)$ between pairs of nodes $i$ and $j$ with a constraint $m(i,j)$, and the following objective function can be used to measure how all pairwise constraints are respected:
\begin{equation}\label{eqn09}
    \mathcal{L}_c=\frac{1}{|\mathcal{M}|}\sum_{m(i,j)\in\mathcal{M}}\exp^{-{||\mathbf{h}_i-\mathbf{h}_j||_2}}
\end{equation}

where $||.||_2=\sqrt{\sum_{i=1}^d|._i|^2}$ is the Euclidean Norm to calculate the Euclidean distance, $\mathcal{M}$ is the set of heterophily constraints, and $d$ is the dimension of representations. The exponential loss~\citep{friedman2001elements} is used to penalize the distance between pairwise constrained nodes.

\subsubsection{Objective function.}
The \textit{constraint-based learning} model mainly contains three components contrastive graph learning framework, graph diffusion convolution, and constraints optimization. It aims to model both the homophily information of the graph structure and pairwise heterophily constraints. 
The objective function of \textit{constraint-based learning} model can be formulated as the combination of unsupervised graph learning $\mathcal{L}_g$ and constraints optimization $\mathcal{L}_c$: 
\begin{equation}
\begin{aligned}
    \mathcal{L}=\mathcal{L}_g + \lambda\cdot\mathcal{L}_c
\end{aligned}
\label{loss_g}
\end{equation}
where the parameter $\lambda\in(0,1)$ controls the importance of the heterophily constraints loss.

\subsection{Part 2: \textit{Constraint-based Binning}}

After obtaining node-level latent representations $\mathbf{H}$ from our \textit{Constraint-based Learning} model, we can directly employ clustering algorithms (K-Means) on these features to obtain final bins. However, two challenges remain to be addressed. First, the number of constraints with respect to each contig vary a lot (i,e., from 0 to dozens); Second, the numbers of contigs in distinct bins/species in metagenomic samples are imbalanced (\emph{i.e.}, varying from a couple to hundreds). 

To address the above challenges, we introduce a constraints-based label annotation strategy (\textit{Constraint-based Binning}) instead of naive clustering algorithms which over rely on the initial centroids. The \textit{Binning} method comprises of two steps, initializing labels for constraints and GCN-based label annotation (see Figure~\ref{fig2} part 2).

\subsubsection{Initializing labels for constraints.} 
Note that not all contigs have pairwise constraints and contigs without any prior constraints are usually more challenging to bin correctly. We first run a clustering algorithm (K-Means in this work) on the representations of nodes with at least one constraint to obtain initial labels, $Y_m=\{y^m_1,y^m_2,...,y^m_k\}$. 
Ideally, the number of bins $K$ should be equal to the number of contigs that contain the same single-copy marker gene. As there are errors in both assemblies and alignments, we need a robust estimate and so, we set $K$ as the third quartile value of the counts of contigs containing each of the marker genes.

\subsubsection{GCN-based label annotation.} After obtaining initial labels for nodes with at least one constraint, we treat the binning as a semi-supervised label annotation task and construct a two-layer graph diffusion convolutional operator to annotate unlabelled contigs (\emph{i.e.}, nodes without any prior constraints). 
Similar to the layer-wise propagation rule defined in Equation~\ref{eqn06}, labels for all node are obtained simultaneously via $\mathbf{Z}=\sigma(\mathbf{T}_S\cdot\mathbf{H}\Theta^{'})=\sigma(\mathbf{T}_S\cdot(\sigma(\mathbf{T}_S\cdot\mathbf{X}\Theta))\Theta^{'})$. We evaluate the cross-entropy error over all contigs with initial labels and use back propagation for optimization.

\begin{equation}
\begin{aligned}
    \mathcal{L}=-\sum_{l\in Y_m}\sum_{k=1}^KY_{lk}\ln\mathbf{Z}_{lk}
\end{aligned}
\label{loss_c}
\end{equation} 

where $Y_m$ is the set of constrained nodes with initialized labels, $K$ is the number of clusters. The pseudocode for RepBin is shown in Algorithm 1.


\begin{algorithm}[tb]
\footnotesize
\caption{The RepBin algorithm}
\label{alg:algorithm}
\textbf{Input}: Assembly Graph $G=(V,E)$, constraints $\mathcal{M}$, num. of bins $k$, initial parameters (\emph{e.g.}, $\lambda$ and $\alpha$);\\
\textbf{Output}: Bins $Y$;
\begin{algorithmic}[1] 
\STATE \textbf{Model Construction}:
\STATE Calculate graph diffusion convolution $\mathbf{T}_S$;
\STATE Construct \textit{Constraint-based Learning} model:
\STATE \ \ \ \ \ \ Corrupt negative graph $\mathcal{C}(X,A)$;
\STATE \ \ \ \ \ \ Learn postive and negative graphs $H$ and $\tilde{H}$;
\STATE \ \ \ \ \ \ Capture global patterns by Readout function $R$;
\STATE \ \ \ \ \ \ Maximize the mutual info. by a discriminator $D$;
\STATE Obtain latent features $\mathbf{H}$;
\STATE Initialize labels for constrained contigs $Y_m$;
\STATE GCN-based label annotation to obtain final bins $Y$;
\STATE \textbf{Optimization:}
\STATE Initialize Embeddings $\mathbf{X}$ with initial features;
\STATE Randomly sample negative graph by corruption $\mathcal{C}$;
\STATE Optimize loss (\ref{loss_g}) via gradient descent;
\STATE Optimize loss (\ref{loss_c}) via gradient descent;
\STATE \textbf{return} Bins $Y$;
\end{algorithmic}
\end{algorithm}

\section{Experiments}

\noindent\textbf{Datasets.} 
Three simulated and two real-world datasets are used in our experiments. Their detailed statistics are given in Table~\ref{tab:dataset}. \textbf{Sim-5G, Sim-10G, and Sim-20G} are simulated based on the species found in the simMC+ dataset~\cite{MaxBin2014}. \textbf{Sharon} is a preborn infant gut metagenome dataset~\cite{Sharon2013}, and the corresponding NCBI accession number is \textit{SRA052203}\footnote{https://www.ncbi.nlm.nih.gov/sra/?term=SRA052203}. 
\textbf{CAMI} is a publicly available dataset extracted from the first CAMI challenge with low complexity of microbiomes~\cite{CAMI2017}\footnote{https://data.cami-challenge.org/participate}. 
All the data sets are assembled using metaSPAdes~\cite{nurk2017metaspades}.

\begin{table}[th]
  \centering
  \begin{tabular}{cccccc}
    \hline
    \hline
    Datasets & Nodes & Edges & $\mathcal{M}$ & Bins & $H$\\
    \hline
    Sim-5G & 519 & 2,488 & 810 & 5 & 0.97 \\
    Sim-10G & 920 & 4,210 & 2,448 & 10 & 0.96 \\
    Sim-20G & 1,452 & 6,531 & 10,734 & 20 & 0.91 \\
    \hline
    CAMI & 5,814 & 23,257 & 21,362 & 19 & 0.89 \\
    Sharon & 20,743 & 102,918 & 2,988 & 12 & 0.95 \\
    \hline
    \hline
  \end{tabular}
  \caption{Statistics of datasets. $\mathcal{M}$ is the number of constraints and $H$ is the node-homophily of the assembly graph.}
  \label{tab:dataset}
\end{table}

\noindent\textbf{Baselines.} 
We compare RepBin with 4 unsupervised GNN models, 4 graph clustering methods, and 8 binning tools. 
\textbf{GNNs:} GraphSAGE~\cite{Hamilton2017graphsage}, GAT~\cite{velickovic2018gat}, DGI~\cite{velickovic2018DGI}, and VGAE~\cite{kipf2016VGAE}. 
\textbf{GCs:} O2MAC~\cite{Fan2020O2MAC}, AGC~\cite{Zhang2019AGC}, Constrainted Spectral Clustering (CSC)~\cite{wang2014constrained}, and Deep Constrained Clustering (DCC)~\cite{zhang2019dcc}. 
\textbf{Binnings:} MetaWatt~\cite{MetaWatt2012}, CONCOCT~\cite{alneberg2014binning}, MaxBin2~\cite{Wu2015MaxBin2}, BusyBeeWeb~\cite{BusyBeeWeb2017}, MetaBAT2~\cite{kang2019metabat}, SolidBin~\cite{Wang2019SolidBin}, VAMB~\cite{nissen2021VAMB}, and GraphBin~\cite{vijini2020graphbin}.

\textbf{Metrics and Experimental Settings.} 
The entire procedure is repeated five times to avoid any experimental biases. Mean and standard deviation values of the binning evaluation metrics are reported. 
For the simulated and CAMI datasets, we map the contigs to the reference genomes to obtain the ground truth species. For the Sharon dataset, we map the contigs to the annotated results\footnote{https://ggkbase.berkeley.edu/carrol/organisms} to obtain the ground truth species.
We use the F1, ARI, and NMI as the evaluation metrics for machine learning-based baselines to evaluate the performance of \textit{Constrain-based Learning} (referred as RepBin-\textit{Learning}).
We use the Precision, Recall, and F1 scores to evaluate the performance of binning contigs.
As existing binning tools typically discard many short contigs and thus cannot bin all contigs, ARI and NMI are not included to avoid possible biases towards binned contigs. We also report the number of bins identified by binning tools.

For RepBin, we use the adjacency matrix as the initial features for nodes. 
The representation dimensions are all empirically set to be 32. 
For all baseline methods, we optimize their models with different parameters and report the best performance scores. 
The RepBin model is freely available\footnote{https://github.com/xuehansheng/RepBin}.

\begin{table*}[t]
\footnotesize
\centering
\begin{tabular}{c|c|c|c|c|c|c|c|c|c|c}
\hline
\hline
  \multicolumn{2}{c|}{\multirow{2}{*}{\textbf{Methods}}} & \multicolumn{3}{c|}{\textbf{Sim-5G}} & \multicolumn{3}{c|}{\textbf{Sim-10G}} & \multicolumn{3}{c}{\textbf{Sim-20G}} \\
  \cline{3-11}
  \multicolumn{2}{c|}{} & \textbf{F1} & \textbf{ARI} & \textbf{NMI} & \textbf{F1} & \textbf{ARI} & \textbf{NMI} & \textbf{F1} & \textbf{ARI} & \textbf{NMI} \\
 \hline
 \multirow{4}{*}{\textbf{GNNs}} & \textbf{GSAGE} & 88.0$\pm$0.6 & 72.9$\pm$0.9 & 81.6$\pm$0.8 & \underline{76.6$\pm$0.2} & \underline{59.3$\pm$0.7} & \underline{75.9$\pm$0.6} & 77.4$\pm$0.6 & 61.5$\pm$1.8 & 81.5$\pm$0.3\\
  & \textbf{GAT} & \underline{94.5$\pm$1.6} & \underline{86.9$\pm$1.7} & \underline{87.7$\pm$0.7} & 73.7$\pm$0.5 & 54.0$\pm$2.6 & 74.3$\pm$0.4 & \underline{79.8$\pm$1.0} & \underline{65.4$\pm$1.3} & \underline{83.6$\pm$0.7} \\
  & \textbf{DGI} & 79.9$\pm$3.4 & 54.6$\pm$6.8 & 68.2$\pm$3.3 & 68.1$\pm$1.9 & 39.3$\pm$2.4 & 61.8$\pm$2.4 & 63.9$\pm$2.5 & 40.1$\pm$2.3 & 65.8$\pm$2.5\\
  & \textbf{VGAE} & 85.7$\pm$1.4 & 70.1$\pm$2.6 & 80.1$\pm$3.2 & 71.4$\pm$1.9 & 46.0$\pm$2.6 & 68.9$\pm$1.1 & 72.2$\pm$1.7 & 54.8$\pm$2.1 & 75.9$\pm$1.2\\
 \hline
 \multicolumn{2}{c|}{\textbf{RepBin-\textit{Learning}}} & \cellcolor{gray!15}99.8$\pm$0.0 & \cellcolor{gray!15}99.4$\pm$0.0 & \cellcolor{gray!15}99.2$\pm$0.0 & \cellcolor{gray!15}97.1$\pm$0.8 & \cellcolor{gray!15}95.4$\pm$1.3 & \cellcolor{gray!15}96.3$\pm$0.8 & \cellcolor{gray!15}91.8$\pm$0.5 & \cellcolor{gray!15}84.8$\pm$0.6 & \cellcolor{gray!15}92.1$\pm$0.2 \\
 \hline
 \multirow{4}{*}{\textbf{GCs}} & \textbf{O2MAC} & 74.9$\pm$3.5 & 63.6$\pm$2.1 & 72.5$\pm$3.1 & 65.8$\pm$1.4 & 53.5$\pm$2.1 & 69.1$\pm$1.7 & 61.0$\pm$3.4 & 52.0$\pm$2.7 & 71.1$\pm$1.8 \\
  & \textbf{AGC} & 80.9$\pm$0.4 & 92.7$\pm$0.8 & 90.4$\pm$0.5 & 78.3$\pm$0.3 & \underline{87.9$\pm$0.3} & \underline{89.6$\pm$0.7} & 67.0$\pm$0.2 & \underline{75.9$\pm$1.1} & 82.0$\pm$0.5 \\
  & \textbf{CSC} & \underline{96.7$\pm$0.0} & 87.9$\pm$0.0 & \underline{91.5$\pm$0.0} & 90.9$\pm$1.3 & 77.9$\pm$4.2 & 85.1$\pm$1.7 & \underline{83.0$\pm$1.4} & 63.2$\pm$4.3 & \underline{83.1$\pm$1.1} \\
  & \textbf{DCC} & 90.9$\pm$0.0 & \underline{94.0$\pm$1.0} & 88.6$\pm$1.1 & \underline{92.1$\pm$2.9} & 83.3$\pm$3.0 & 77.3$\pm$2.3 & 82.1$\pm$1.9 & 65.3$\pm$2.8 & 75.1$\pm$3.3 \\
 \hline
 \multicolumn{2}{c|}{\textbf{RepBin}} & \cellcolor{gray!15}99.7$\pm$0.1 & \cellcolor{gray!15}99.0$\pm$0.2 & \cellcolor{gray!15}98.5$\pm$0.0 & \cellcolor{gray!15}99.4$\pm$0.0 & \cellcolor{gray!15}99.1$\pm$0.1 & \cellcolor{gray!15}98.8$\pm$0.3 & \cellcolor{gray!15}97.2$\pm$0.6 & \cellcolor{gray!15}94.3$\pm$0.8 & \cellcolor{gray!15}95.7$\pm$0.6 \\
 \hline
\hline
\end{tabular}
\caption{The results of RepBin and machine leaning-based baselines on three simulated datasets for contigs binning (\%).}
\label{tab:exp0}
\end{table*}

\begin{figure*}[t]
\centering
\includegraphics[width=1.75\columnwidth]{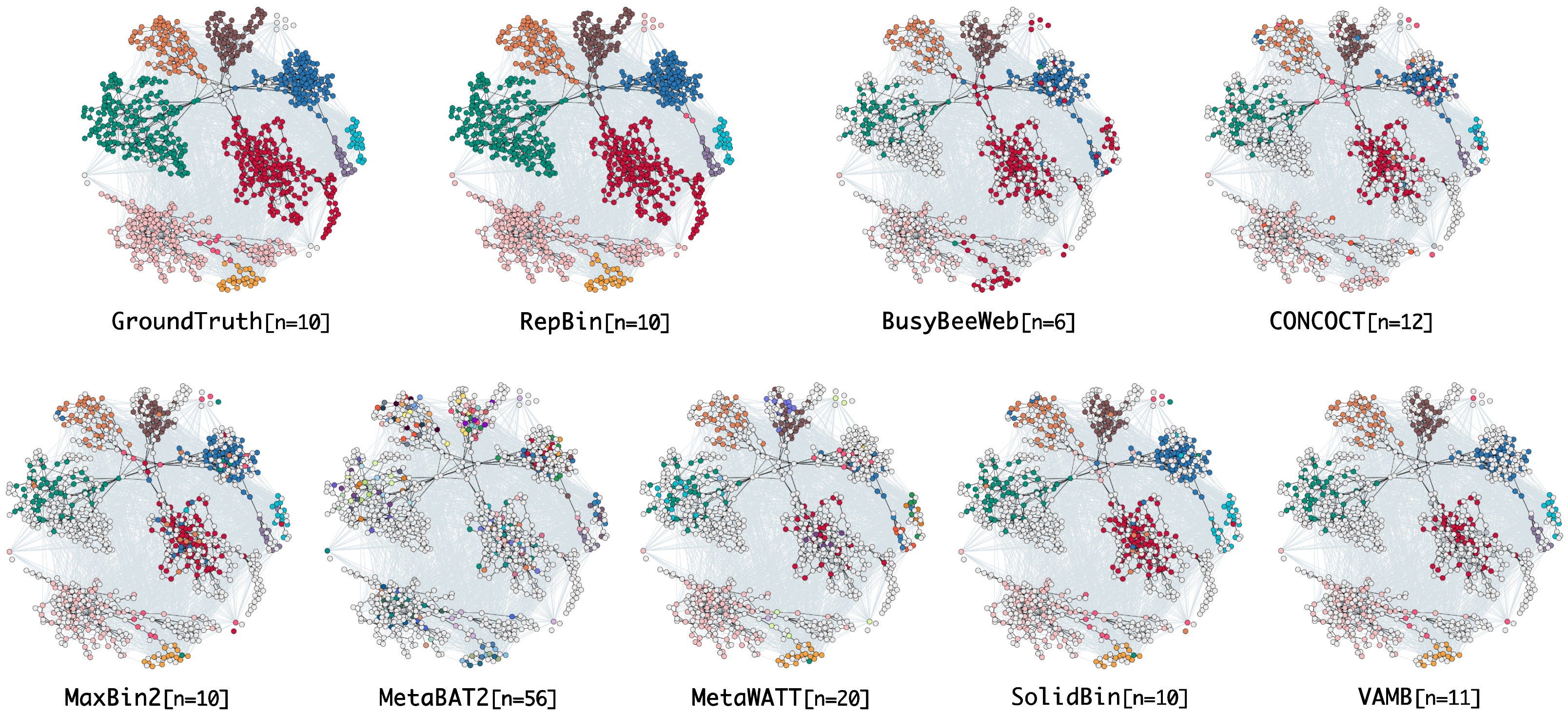}
\caption{Visualization of the Sim-10G assembly graph with ground truth and different binning results. 
}
\label{visualization}
\end{figure*}

\subsection{Benchmarking against Machine Learning Models}

Table~\ref{tab:exp0} shows that Both RepBin-\textit{Learnig} and RepBin significantly outperform machine learning-based baselines, which achieve the highest scores on all metrics and all three datasets. For Sim-20G, the metric scores obtained by RepBin-\textit{Learning} are 91.8$\%$ for F1, 84.8$\%$ for ARI, and 92.1$\%$ for NMI respectively which is significantly higher than the highest scores achieved by GNNs (79.8$\%$ for F1, 65.4$\%$ for ARI, and 83.6$\%$ for NMI). Both GNNs and RepBin-\textit{Learning} run K-Means algorithm to cluster nodes after obtaining representations for nodes. The gap between GNNs and RepBin-\textit{Learning} demonstrates the superiority of our model in learning structure of graph and integrating prior constraints. 
Because of the label imbalance, graph clustering (GCs) methods also achieve lower metric scores than RepBin. The results of two constraint-based clustering algorithms are also inferior to RepBin, because these methods cannot capture structural information of the graph well.

\begin{table*}[t]
\footnotesize
\centering
\begin{tabular}{c|c|c|c|c|c|c|c|c|c}
\hline
\hline
 \multicolumn{2}{c|}{\multirow{2}{*}{\textbf{Datasets}}} & \multirow{2}{*}{\textbf{MetaWatt}} & \textbf{CON} & \multirow{2}{*}{\textbf{MaxBin2}} & \textbf{BusyBee} & \multirow{2}{*}{\textbf{MetaBAT2}} & \multirow{2}{*}{\textbf{SolidBin}} & \multirow{2}{*}{\textbf{VAMB}} & \multirow{2}{*}{\textbf{RepBin}} \\
 \multicolumn{2}{c|}{} &  & \textbf{COCT} &  & \textbf{Web} & & & \\
 \hline
\multirow{4}{*}{\textbf{Sim-5G}} & Precision & \cellcolor{gray!15}\underline{100.00} & 91.60 & 91.13 & 86.57 & \cellcolor{gray!15}\underline{100.00} & 90.00 & \cellcolor{gray!15}\underline{100.00$\pm$0.00} & 99.69$\pm$0.10 \\
    & Recall & 24.59 & 40.50 & 46.69 & \underline{49.79} & 6.61 & 46.49 & 33.92$\pm$0.90 &  \cellcolor{gray!15}99.69$\pm$0.10 \\
    & F1 & 39.47 & 56.16 & 61.75 & \underline{63.22} & 12.40 & 61.31 & 50.66$\pm$1.02 & \cellcolor{gray!15}99.69$\pm$0.10 \\
    & Pred. bins & 12 & 7 & 5 & 4 & 34 & 5 & 6 & 5 \\
\hline
\multirow{4}{*}{\textbf{Sim-10G}} & Precision & 99.29 & 86.99 & 89.43 & 84.47 & \cellcolor{gray!15}\underline{100.00} & 91.58 & \underline{99.93$\pm$0.15} & 99.20$\pm$0.00 \\
    & Recall & 26.13 & 39.72 & 40.30 & \underline{45.53} & 6.39 & 41.70 & 33.80$\pm$0.20 & \cellcolor{gray!15}99.55$\pm$0.08 \\
    & F1 & 41.38 & 54.54 & 55.56 & \underline{59.17} & 12.01 & 57.30 & 50.51$\pm$0.23 & \cellcolor{gray!15}99.37$\pm$0.04 \\
    & Pred. bins & 20 & 12 & 10 & 6 & 56 & 10 & 11 & 10 \\
\hline
\multirow{4}{*}{\textbf{Sim-20G}} & Precision & 96.85 & 84.02 & 88.25 & 77.39 & 96.77 & 96.51 & \cellcolor{gray!15}\underline{99.35$\pm$0.10} & 97.31$\pm$0.31 \\
    & Recall & 32.01 & 42.27 & 41.69 & 44.51 & 7.73 & \underline{85.04} & 36.88$\pm$0.60 & \cellcolor{gray!15}96.98$\pm$0.69 \\
    & F1 & 48.12 & 56.24 & 56.63 & 56.52 & 14.32 & \underline{90.41} & 53.79$\pm$0.64 & \cellcolor{gray!15}97.15$\pm$0.61 \\
    & Pred. bins & 33 & 22 & 21 & 12 & 88 & 20 & 22 & 20 \\
\hline
\multirow{4}{*}{\textbf{CAMI}} & Precision & 86.29 & 92.83 & 85.33 & 77.47 & 95.41 & 80.65 & \cellcolor{gray!15}\underline{98.82$\pm$0.41} & 83.67$\pm$0.93 \\
    & Recall & 37.12 & 43.65 & 39.69 & \underline{53.15} & 1.51 & 44.86 & 31.90$\pm$0.92 & \cellcolor{gray!15}92.84$\pm$0.78 \\
    & F1 & 51.91 & 59.38 & 54.18 & \underline{63.05} & 2.97 & 57.66 & 48.23$\pm$1.06 & \cellcolor{gray!15}87.41$\pm$0.60 \\
    & Pred. bins & 65 & 30 & 22 & 16 & 101 & 20 & 21 & 17 \\
\hline
\multirow{4}{*}{\textbf{Sharon}} & Precision & 79.08 & 74.03 & 82.33 & 68.68 & 76.89 & 70.04 & \cellcolor{gray!15}\underline{97.41$\pm$0.21} & 73.60$\pm$0.72 \\
    & Recall & 18.91 & 24.58 & 25.74 & \underline{43.82} & 1.72 & 25.82 & 30.23$\pm$2.51 & \cellcolor{gray!15}76.59$\pm$0.79 \\
    & F1 & 30.52 & 36.91 & 40.37 & \underline{53.50} & 3.36 & 37.73 & 46.10$\pm$3.01 & \cellcolor{gray!15}74.97$\pm$0.16 \\
    & Pred. bins & 39 & 48 & 16 & 5 & 34 & 9 & 9 & 11 \\
\hline
\multicolumn{2}{c|}{} & \multicolumn{7}{c|}{\textbf{Stand-alone Binning Tools + GraphBin}} & \\
\hline
\multirow{4}{*}{\textbf{Sim-20G}} & Precision & 98.02 & 92.96 & 96.77 & 90.27 & 96.77 & 96.51 & \cellcolor{gray!15}\underline{98.15$\pm$0.04} & 97.31$\pm$0.31 \\
    & Recall & 68.71 & 81.86 & 83.96 & \underline{91.84} & 7.73 & 85.04 & 86.35$\pm$2.21 & \cellcolor{gray!15}96.98$\pm$0.69 \\
    & F1 & 80.79 & 87.06 & 89.91 & 91.05 & 14.32 & 90.41 &  \underline{91.86$\pm$1.25} & \cellcolor{gray!15}97.15$\pm$0.61 \\
    & Pred. bins & 33 & 22 & 21 & 12 & 88 & 20 & 22 & 20 \\
\hline
\multirow{4}{*}{\textbf{CAMI}} & Precision & 91.88 & \cellcolor{gray!15}\underline{96.44} & 89.18 & 85.16 & 92.44 & 86.86 & 95.05$\pm$0.26 & 83.67$\pm$0.93 \\
    & Recall & 60.12 & 79.43 & 76.82 & \underline{88.17} & 36.77 & 82.15 & 78.94$\pm$1.47 & \cellcolor{gray!15}92.84$\pm$0.78 \\
    & F1 & 72.68 & \underline{87.11} & 82.54 & 86.85 & 52.60 & 84.44 & 86.24$\pm$0.91 & \cellcolor{gray!15}87.41$\pm$0.60 \\
    & Pred. bins & 65 & 30 & 22 & 16 & 101 & 20 & 21 & 17 \\
\hline
\hline
\end{tabular}
\caption{The experimental metrics of RepBin and baselines on five datasets for binning contigs (\%).}
\label{tab:exp1}
\end{table*}

\subsection{Benchmarking against Metagenomic Binning Tools}
Table~\ref{tab:exp1} shows the results obtained by RepBin and stand-alone baselines on all five datasets. We also compare RepBin with refined binning results of baselines with GraphBin on Sim-20G and CAMI datasets. 

Table~\ref{tab:exp1} shows that RepBin significantly outperforms baselines including GraphBin refinements. RepBin achieves the highest Recall and F1 score on all simulated and real-world datasets. Take Sim-20G for example, the highest Recall and F1 score achieved among all baselines is SolidBin, 85.04$\%$ and 90.41$\%$, which is much lower than the scores achieved by RepBin (96.98$\%$ for Recall and 97.15$\%$ for F1). The Precision score achieved by RepBin is 97.31$\%$, which is slightly lower than VAMB (99.35$\%$) and significantly higher than other baselines. In the publicly-available CAMI dataset, RepBin also achieves the highest Recall and F1 score (92.84$\%$ and 87.41$\%$) against other baselines (the second highest score achieved by BusyBeeWeb, (53.15$\%$ for Recall and 63.05 $\%$ for F1). The significant improvement achieved by RepBin on the F1 score indicates that RepBin can tackle the label imbalance problem well and accurately bin as many contigs as possible. Moreover, Table~\ref{tab:exp1} shows that RepBin also achieves better performance against the refined binning results of baselines+GraphBin. 
GraphBin is not a stand-alone binning tool. We have to first run one existing binning tool (e.g., MetaWatt) to derive the initial binning results and then run GraphBin to improve the initial binning results. 
Although GraphBin improves the binning results on all stand-alone baselines, the final evaluation matrices are still inferior to RepBin.
\textbf{Visualization.} To better understand the binning results, we use python-iGraph package to visualize the Sim-10G assembly graph with ground truth and binning results from distinct stand-alone binning tools (see Figure~\ref{visualization}). 
Different colors denote distinct species and grey nodes indicate the nodes that are not identified or are discarded. 
Black edges represent homophily edges in the assembly graph and grey edges are heterophily constraints. 
RepBin achieves the most consistent labels with respect to the ground truth while other baselines suffering from missing or incorrect labels.



\subsection{Ablation Study}

\begin{figure}[t]
\centering
\includegraphics[width=0.975\columnwidth]{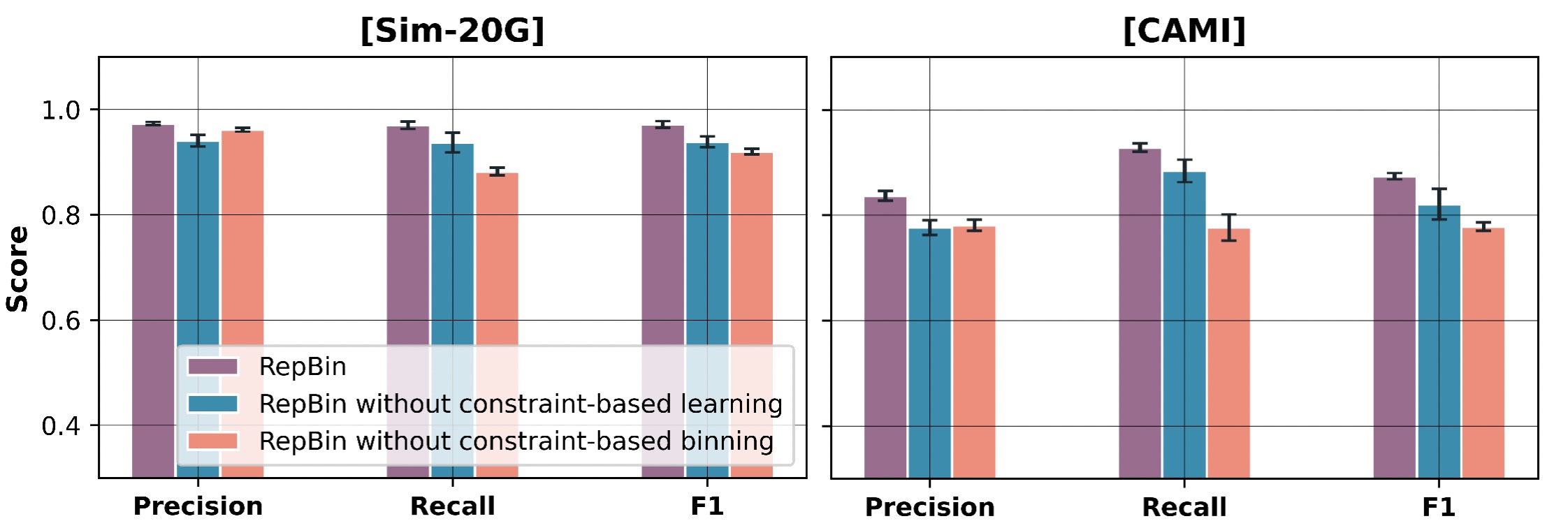}
\caption{Results of RepBin and its variants on Sim-20G and CAMI datasets.}
\label{comp}
\end{figure}

\textbf{Effects of Constraints in Learning and Binning.} 
To study the effectiveness of incorporating constraints of our model, we conduct an ablation study by examining variants of RepBin. 
\textit{RepBin without constraint-based learning} represents RepBin discards the information of constraints in the Learning part and only captures the structure of the assembly graph. \textit{RepBin without constraint-based binning} denotes RepBin without constraint-based binning part (refer to the \textit{Constraint-based Learning} model) and uses K-Means algorithm to obtain final binning results.
Figure~\ref{comp} shows that modelling constraints in RepBin improves metagenomic binning. For CAMI, the F1 values achieved by RepBin, \textit{RepBin without constraint-based learning}, \textit{RepBin without constraint-based binning} are 87.41$\%$, 82.10$\%$, and 77.84$\%$ respectively, which demonstrate the effectiveness of constraint-based learning and binning.

\begin{figure}[t]
\centering
\includegraphics[width=0.95\columnwidth]{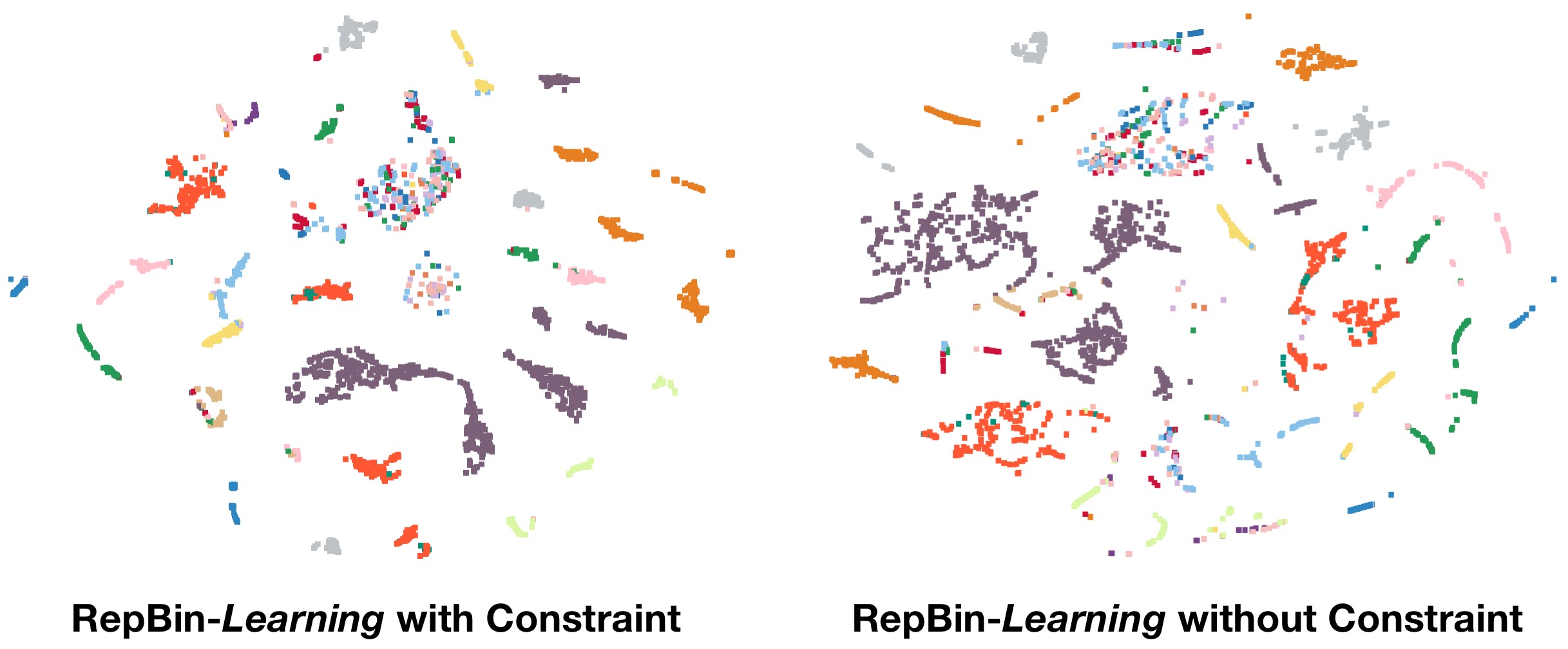}
\caption{Visualization of representations on CAMI.}
\label{embed}
\end{figure}

Besides, we also use t-SNE~\cite{2008tsne} to visualize the latent embeddings of \textit{RepBin}. Figure~\ref{embed} shows the 2D-visualization of the embeddings from the CAMI dataset, where nodes with distinct colors represent different species. Visually, \textit{RepBin-Learning with Constraint} gives the best separation between different clusters comparing with \textit{RepBin-Learning without Constraint}.

\subsubsection{Parameters Analysis.} 
\begin{figure}[t]
\centering
\includegraphics[width=0.975\columnwidth]{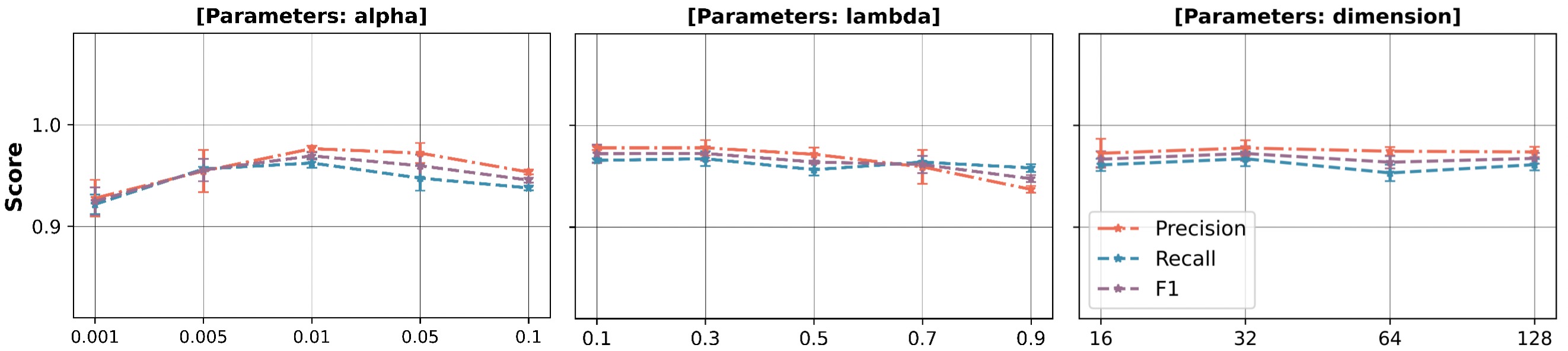}
\caption{Parameters analysis of the RepBin model.}
\label{params}
\end{figure}

The parameter $\alpha$, varying from 0.001 to 0.1, is used in diffusion to control the probability of teleporting to another state. 
The parameter $\lambda$, varying from 0.1 to 0.9, is used to balance the importance between the graph and constraints in the loss function of the \textit{Constraint-based Learning} model.
The parameter $d$, varying in $\{16, 32, 64, 128\}$, represents the dimension of the RepBin-\textit{Learning}. 
Figure~\ref{params} shows that the performance of RepBin is not sensitive to the changes in above parameters.

\subsubsection{Training \& Running Analysis.} 
\begin{figure}[t]
\centering
\includegraphics[width=0.975\columnwidth]{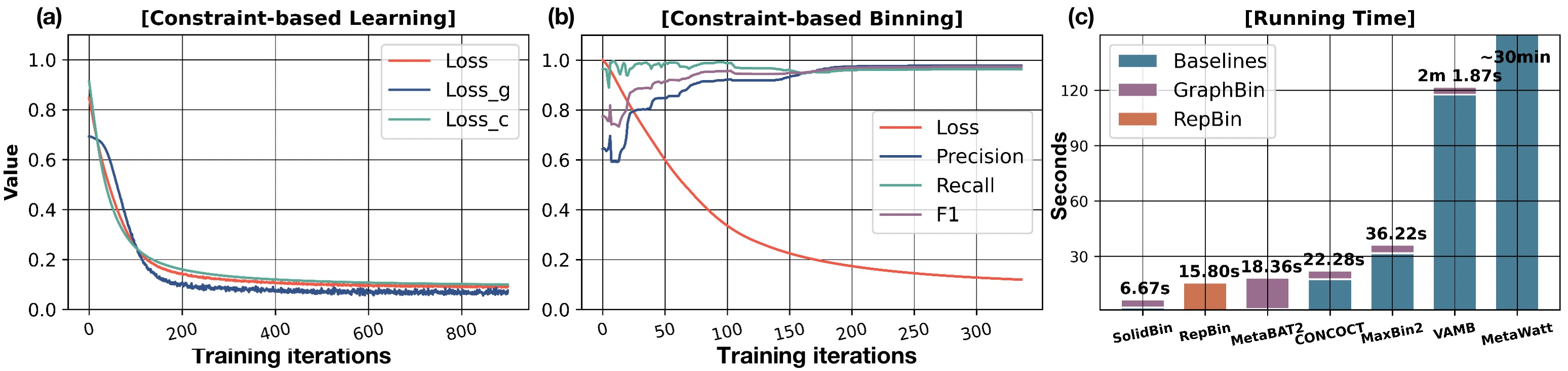}
\caption{The training loss and running time of RepBin.}
\label{training}
\end{figure}

Figure~\ref{training} (a) and (b) show the process of optimizing Equation~\ref{eqn06} and the changes of evaluation metrics in the \textit{Constraint-based Binning} model of the RepBin. Figure~\ref{training} (c) shows the running time of RepBin against other baselines on Sim-20G. RepBin is the second fastest binning tool and only slightly slower than SolidBin. 


\section{Conclusion}
To learn both the graph structure and prior heterophily information (represented as pairwise constraints), we propose a constraint-based graph representation learning model, \textit{Constraint-based Learning}. It is a contrastive graph learning framework that uses a diffusion graph convolutional operator to model graph structure and respects prior constraints by adding a penalty with respect to constrained nodes. 
We also propose a \textit{Constraint-based Binning} model which groups constrained nodes using k-means algorithm and then use a semi-supervised GCN model to annotate unlabeled nodes, which is robust to imbalance in the node constraints and the bin sizes. 
The proposed \textit{RepBin} is implemented to solve the real-world task of metagenomic binning. Extensive experiments demonstrate the superiority of our proposed model in contigs binning.

\bibliography{aaai22}

\section{Appendix}

\subsection{A: Distributions of Bin Sizes and Node Constraints}
\begin{figure}[h]
\centering
\includegraphics[width=0.99\columnwidth]{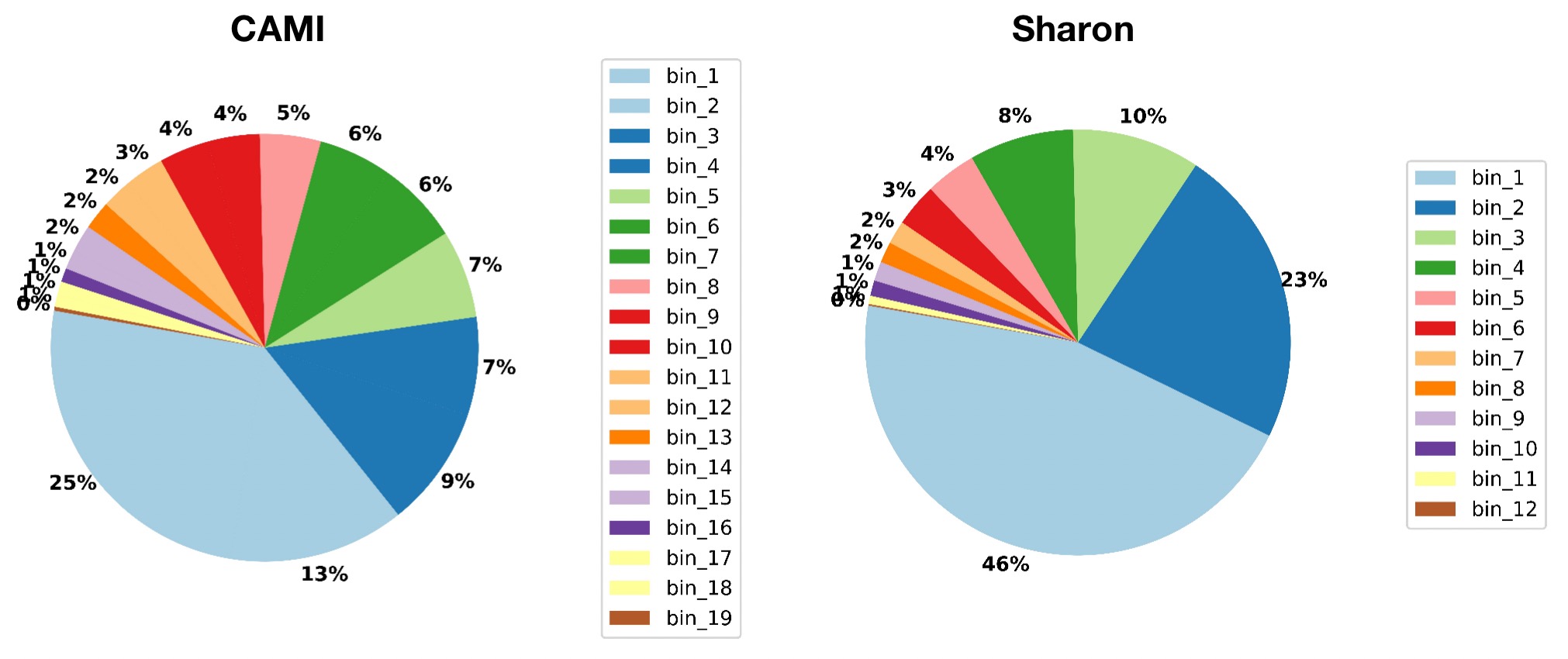}
\caption{The distribution of bin sizes for CAMI and Sharon.}
\label{figs1}
\end{figure}

Figure~\ref{figs1} shows the pie chart of the varying bin sizes for two publicly-available datasets, CAMI and Sharon. In CAMI, the three largest bins contain nearly 50\% contigs. Similary, in Sharon, the largest bin alone occupies 46\% contigs. The imbalance sizes of bins makes it more difficult for metagenomic binning.

\begin{figure}[h]
\centering
\includegraphics[width=0.99\columnwidth]{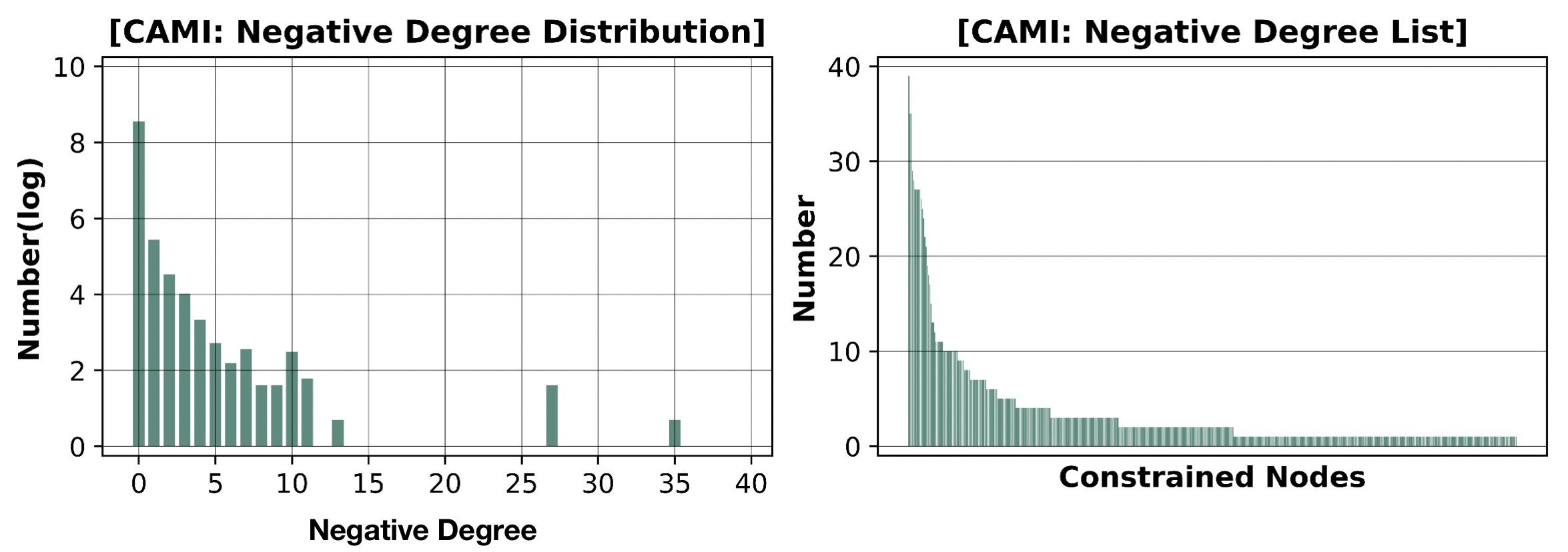}
\caption{The negative degree distribution and list of constrained nodes.}
\label{figs2}
\end{figure}

Figure~\ref{figs2} shows the distribution and list of the number of heterophily constraints that a node participates. In CAMI, while ~90\% of the nodes are not associated with any heterophily constraints and 4\% of the constrained nodes only have one negative degree (the left figure). 
There also exists a node with 39 constraints (the right figure). This imbalance of constrained nodes is also observed in other metagenomic datasets.

\subsection{B Evaluation Metrics}
We use Precision, Recall, and F1 as evaluation metrics for the task of contigs binning. The confusion matrix $A\in\mathcal{R}^{K\times S}$ is widely to represent the binning results where $K$ is the number of bins predicted by binning tools and $S$ denotes the number of true species in the ground truth. The element $A_{k,s}$ denotes the number of contigs are classified into the bin $k$ but actually belong to the specie $s$. $N$ is the total number of contigs binned by this tool and $N_{u}$ denotes the number of contigs are not classified or discarded by this binning tool. The detailed evaluation metrics are described as follow:

\begin{equation}
\begin{aligned}
  &  Precision = \frac{\sum_k\max_s{A_{k,s}}}{\sum_k\sum_sA_{k,s}}, \\
  &  Recall = \frac{\sum_s{\max_k{A_{k,s}}}}{\sum_k\sum_sA_{k,s}+N_{u}}, \\
  &  F1 = 2\times\frac{Precision\times Recall}{Precision + Recall}.
\end{aligned}
\end{equation}

In the results, we also show the number of identified by different binning tools, marked as $\mathrm{Pred. bins}$. 
Binning tools with higher performance generally yield more accurate estimation on $\mathrm{Pred. bins}$ compared with ground truth. Both $\mathrm{Pred. bins}$ higher or lower than ground truth  are not good.

In addition, we use F1, ARI, and NMI to evaluate the performance of RepBin and baselines on graph clustering. The euqation of ARI and NMI are listed as follow:
\begin{equation}
    ARI = \frac{\sum_{k,s}\tbinom{A_{k,s}}{2}-\tfrac{\sum_k\tbinom{A_{k,.}}{2}\sum_s\tbinom{A_{.,s}}{2}}{\tbinom{N}{2}}}{\tfrac{1}{2}(\sum_k\tbinom{A_{k,.}}{2}+\sum_s\tbinom{A_{.,s}}{2})-\tfrac{\sum_k\tbinom{A_{k,.}}{2}\sum_s\tbinom{A_{.,s}}{2}}{\tbinom{N}{2}}} \\
\end{equation}

\begin{equation}
    NMI = \frac{2\times\sum_{k,s}A_{k,s}\log(\tfrac{N\times A_{k,s}}{A_{k,.}\times A_{.,s}})}{\sum_sA_{.,s}\log\tfrac{A_{.,s}}{N}+\sum_kA_{k,.}\log\tfrac{A_{k,.}}{N}}
\end{equation}

\subsection{C Detailed Information of Baselines}
\subsubsection{Graph neural networks (GNNs):}\ \\
%
\noindent\textbf{GraphSAGE.} It is an inductive graph representation learning model by iteratively sampling and aggregating messages from neighbors and finally generates features for nodes~\cite{Hamilton2017graphsage}.  Available at \url{https://github.com/williamleif/GraphSAGE}.

\noindent\textbf{GAT.} It assigns masked self-attention mechanism for different neighbors to aggregate messages with different weights~\cite{velickovic2018gat}. Available at \url{https://github.com/gordicaleksa/pytorch-GAT}.

\noindent\textbf{DGI.} It is a contrastive graph learning model which maximizes the mutual information between local and graph-level summarized features to capture global structure of the graph~\cite{velickovic2018DGI}. Available at \url{https://github.com/dfdazac/dgi}.

\noindent\textbf{VGAE.} It is a generative graph learning model which learns low-dimensional feature for nodes by minimizing the error between the original and the reconstructed graph~\cite{kipf2016VGAE}. Available at \url{https://github.com/tkipf/gae}.

\subsubsection{Graph clustering:}\ \\
\noindent\textbf{O2MAC.} It contains a generative graph learning model and a self-training clustering algorithm, which can jointly optimize the graph learning loss and clustering loss~\cite{Fan2020O2MAC}. Available at \url{https://github.com/googlebaba/WWW2020-O2MAC}.

\noindent\textbf{AGC.} It proposes an adaptive graph convolutional learning model to capture the high-order structure of the graph and uses a spectral clustering algorithm to cluster nodes~\cite{Zhang2019AGC}. Available at \url{https://github.com/karenlatong/AGC-master}.

\noindent\textbf{CSC.} It encodes constraints as part of a optimization problem and proposes a constrainted spectral clustering method~\cite{wang2014constrained}. Available at \url{https://github.com/peisuke/ConstrainedSpectralClustering}.

\noindent\textbf{DCC.} It combines deep learning with constrained clustering and firstly proposes a deep constrained clustering model~\cite{zhang2019dcc}. Available at \url{https://github.com/blueocean92/deep_constrained_clustering}.

\subsubsection{Metagenomic contigs binning:}\ \\
\noindent\textbf{MetaWatt.} It proposes a multivariable statistics binning model by combining tetranucleotide frequencies and interpolated Markov models~\cite{MetaWatt2012}. Available at \url{https://sourceforge.net/projects/metawatt/}.

\noindent\textbf{CONCOCT.} It uses gaussian mixture models to integrate both sequence composition and coverage across multiple samples to bin contigs into genomes~\cite{alneberg2014binning}. Available at \url{https://github.com/BinPro/CONCOCT}.

\noindent\textbf{MaxBin2.} It calculates the tetranuncleotide frequencies and coverages of the contigs and use an EM algorithm to cluster contigs~\cite{Wu2015MaxBin2}. Available at \url{https://sourceforge.net/projects/maxbin/}.

\noindent\textbf{BusyBeeWeb.} An online bootstrapped supervised contigs binning and annotation web tool~\cite{BusyBeeWeb2017}. Available at \url{https://ccb-microbe.cs.uni-saarland.de/busybee/}.

\noindent\textbf{MetaBAT2.} It adaptively integrates distances of genome abundance and normalized tetra-nucleotide frequency to iteratively partition contigs~\cite{kang2019metabat}. Available at \url{https://bitbucket.org/berkeleylab/metabat}.

\noindent\textbf{SolidBin.} It integrates prior pairwise constraints and clusters metagenomic contigs in a semi-supervised spectral normalized cut manner~\cite{Wang2019SolidBin}. Available at \url{https://github.com/sufforest/SolidBin}.

\noindent\textbf{VAMB.} It uses variational autoencoder models to learn both the sequence coabundance and k-mer distribution and then uses an iterative clustering model to bin contigs~\cite{nissen2021VAMB}. Available at \url{https://github.com/RasmussenLab/vamb}.

\noindent\textbf{GraphBin.} It employs a label propagation algorithm on the assembly graph to better refine the binning results of existing stand-alone tools~\cite{vijini2020graphbin}. Available at \url{https://github.com/Vini2/GraphBin}.

\subsection{D Reproducibility}
\textbf{Running environment.} 
RepBin is implemented in Python 3.6 and Pytorch 1.8 using the Linux server with 6 Intel(R) Core(TM) i7-7800X CPU @ 3.50 GHz, 96GB RAM and 2 NVIDIA TITAN Xp 12 GB. 

\noindent\textbf{Experimental setups.}  
Two types of baselines are employed in our paper, machine learning-based models and metagenomic contigs binning tools. 
We compare RepBin-\textit{Learning} with graph neural network models, graph clustering methods, and constraint-based clustering algorithms. 
RepBin-\textit{Learning} model is an unsupervised graph representation learning model considering pairwise constraints. Thus, we run K-Means to group these nodes into different bins.

For GNNs, we first implement these models in an unsupervised manner to learn the structure of the graph and then run K-Means algorithms on these learned features to obtain final binning results. To evaluate the effectiveness of our proposed RepBin-\textit{Learning} on capturing the information of graph structure, we also compare RepBin-\textit{Learning} with four Deep Graph Clustering models (O2MAC, AGC, CSC, and DCC). O2MAC and AGC do not consider the information of heterophily constraints, and CSC and DCC are constraint-based graph clustering algorithms.
The K-Means algorithm used for clustering and partial evaluation metrics are all from the scikit-learn library~\cite{pedregosa2011scikit}.

\noindent\textbf{Detailed parameters.} 
The initial bin number for different datasets are 5 for Sim-5G, 10 for Sim-10G, 20 for Sim-20G, 19 for CAMI, and 11 for Sharon respectively. 
The dimension of representations is empirically set to 32, and the parameter analysis shows that our model is robust to parameter $h$. 
Parameter $\epsilon$ aims to sparse the dense matrix from graph diffusion convolutions and it can accelerate the training process. Our model is also robust to $\epsilon$ and we simply set this parameter as 1e-4. Parameter $\lambda$ balance the importance of heterophily constraints, and RepBin achieve good performance in the range of [0.1-0.5]. $\alpha$ is the pramater in the graph diffusion convolutional operator. Experiments show that our model achieve good performance varying from 0.005 to 0.05. 
All codes, data and experimental settings of the RepBin model are released at \url{https://github.com/xuehansheng/RepBin}.

\subsection{E Additional Results}
\noindent\textbf{Long contigs.} 
RepBin can bin both short and long contigs. 
We now report the results of RepBin and baselines only considering long contigs ($>$ 1,000 bp) for Sim-20G and the following table shows that RepBin when discarding short contigs still significantly outperforms baselines.

\begin{table}[h]
\footnotesize
\setlength\tabcolsep{4.0pt}
\centering
\begin{tabular}{c|c|c|c|c|c|c|c|c}
\hline
 \multirow{2}{*}{} & Meta & CON & Max & BusyB & Meta & Solid & VA & Rep \\
  & Watt & COCT & Bin2 & eeWeb & BAT2 & Bin & MB & Bin \\
 \hline
 P & 98.0 & 84.0 & 88.3 & 81.1 & 96.8 & 86.8 & 99.4 & 97.2 \\
 R & 60.6 & 86.4 & 85.2 & 89.5 & 15.8 & 85.4 & 74.7 & 95.1 \\
 F1 & 74.9 & 85.2 & 86.7 & 85.1 & 27.2 & 86.1 & 85.3 & 96.1 \\
\hline
 & \multicolumn{7}{c|}{Stand-alone Binning Tools + GraphBin} & \\
\hline
 P & 98.8 & 91.5 & 96.9 & 89.6 & 98.5 & 97.3 & 99.5 & 97.2 \\
 R & 63.4 & 83.0 & 81.8 & 93.1 & 15.1 & 83.9 & 87.6 & 95.1 \\
 F1 & 77.2 & 87.1 & 88.7 & 91.3 & 26.1 & 90.1 & 93.2 & 96.1 \\
\hline
\end{tabular}
\label{tab:01}
\end{table}

\noindent\textbf{Incorrect constraints.} It is possible to introduce incorrect constraints by single-copy marker genes. The following table shows that in the presence of incorrect constraints, RepBin still performs well (using Sim-20G as an example). 

\begin{table}[h]
\footnotesize
\setlength\tabcolsep{5.0pt}
\centering
\begin{tabular}{c|c|c|c|c|c}
\hline
 \multicolumn{2}{c|}{Ratio of incorrect constraints} & 0\% & 10\% & 20\% & 30\% \\
\hline
\multirow{2}{*}{RepBin} & Precision & 96.36 & 92.81 & 90.73 & 88.35 \\
\multirow{2}{*}{-Learning} & Recall & 87.71 & 84.91 & 79.03 & 78.41 \\
    & F1 & 91.81 & 88.68 & 84.48 & 83.07 \\
\hline
\multirow{3}{*}{RepBin} & Precision & 97.31 & 98.14 & 94.65 & 92.42 \\
    & Recall & 96.98 & 95.76 & 95.32 & 93.90 \\
    & F1 & 97.15 & 96.94 & 94.98 & 93.15 \\
\hline
\end{tabular}
\label{tab:02}
\end{table}

\end{document}